# A theoretical estimate on the probability of the formation of a self-avoiding copolymer macromolecule

**Pramod Kumar Mishra**
Email: pkmishrabhu@gmail.com
Department of Physics, DSB Campus, Kumaun University,
Nainital,Uttarakhand (INDIA)

**Abstract**. A lattice model of the directed self-avoiding walk is used to estimate the possibility on the formation of an infinitely long linear semi-flexible copolymer chain. The copolymer chain is assumed to composed of four different types of the monomers. A method of the recursion relations is used to solve the proposed model analytically to show that the probability of the formation of a self-avoiding semi-flexible copolymer chain is independent of the stiffness of the chain. It is a distinct result from our earlier study on the formation of a Gaussian semi-flexible copolymer chain and the Gaussian chain is made up of these four monomers, [P. K. Mishra, J. of Adv. Appl. Sci. Res. 2(4) 1-8 (2020)]. We have also calculated the average number of different types of the bonding in the copolymer chain to show the distinctions in the behaviour of the self-avoiding copolymer chain from the Gaussian polymer chain.

## 1. Introduction

There are well known macromolecules which forms the basis for the formation of the living organisms and the non-living things. These macromolecules are *Lipids*, *Carbohydrates*, *Nucleic Acids* and the *Proteins*; and but the Lipids are the non polymeric macromolecule [1]. These macromolecules are made up of different types of the monomers and these monomers bear a finite volume and therefore there is an excluded volume interaction among the monomers of the real (self-avoiding) chain. Thus, the real copolymer chains are self-avoiding type and the bending in the segments of the chain cost some energy i. e. the bending energy is required to produce kink in the chain. Therefore, we have chosen a self-avoiding walk model for the semi-flexible copolymer chain to estimate the probability of the formation of an infinitely long linear copolymer chain. A directed walk model [2] of the copolymer chain is suitable to solve analytically and so that we may get the exact estimates on the thermodynamics of an infinitely long semi-flexible self-avoiding copolymer chain.

There are several reports on the thermodynamics of an infinitely long homo-polymer chain [2-5] and however, theoretical reports on the thermodynamics of the semi-flexible copolymer chain is limited. Therefore, we have chosen a directed walk model of a self-avoiding copolymer chain to estimate some of the thermo-dynamical properties of the semi-flexible copolymer chain.

The manuscript is organized as follows: a brief discussion is given in the section two on the directed self-avoiding walk model of the semi-flexible copolymer chain; and in the section three we have outlined the method of the calculations and the results obtained. The results are summarized and a discussion on findings of the manuscript has been done in the section four.

## 2. The model and the method

A directed self-avoiding walk model has been extensively used to study the thermodynamics of a flexible and a semi-flexible homo-polymer chain [2-9]. The beauty of the directed walk model is that it can be solved analytically and therefore, it can give us the exact results on the thermodynamics of an infinitely long chain [2-9]. The qualitative findings of directed walk model is identical to isotropic model of the self avoiding polymer chain [5].

We have chosen a partially directed self-avoiding walk model [2-5] on a cubic lattice, where $+x$, $\pm y$ and $+z$ directions are allowed to the walker while enumerating the conformations of a self-avoiding semi-flexible copolymer chain. The semi-flexibility has been taken into account by introducing a Boltzmann weight $k[=Exp(-\beta E_b)]$ of the bending energy ($E_b$) on each bend in the chain. The four monomers of the chain are taken in different possible sequences to understand the role of these sequences on the probability of the formation of an infinitely long self-avoiding semi-flexible copolymer chain. The Boltzmann weights $u[=w^2=Exp(2\beta E_s)]$ and $v[=w^3=Exp(3\beta E_s)]$ corresponds to the energy ($E_s$) which is released in the polymer solution when an *a-t* or a *c-g* bonding between these monomers may occur [10]. A seven monomers long self-avoiding semi-flexible copolymer chain is shown schematically in the figure no. (1)-II for the sake of completeness.

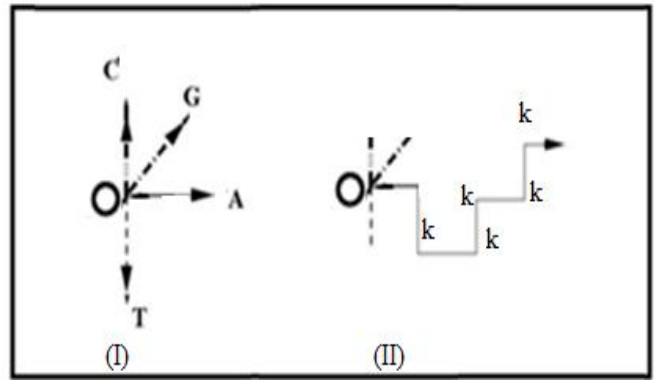

**Figure No. 1:** A partially directed self-avoiding walk of a seven monomers are shown to mimic a conformation of a self-avoiding semi-flexible copolymer chain. The sequence of the monomers (*a*, *t*, *c* and *g*) are shown using figure no. 1 (I). The monomers are arranged in the sequence ***a-t-a-c-a-c-a***. There are six bends in the copolymer chain, and it is shown in figure 1 (II). Thus, the Boltzmann weight of the walk is $g^7 k^6 w^4$.

The grand canonical partition of the semi-flexible copolymer chain is written in general as,

$$G(g, u, v, k) = \sum_{N=1}^{\infty} \sum_{\text{All walks of N monomers}} g^N u^m v^n k^p \qquad (1)$$

Here, g is the minimum value of the step fugacity for each monomer in the chain, $u*g$ and $v*g$ are the effective fugacity of a monomer when there is an *a-t* or a *c-g* bonding in the chain may occur, respectively. There are *m* number of *a-t* bonds, *n* number the *c-g* bonds and *p* number of bends in the copolymer chain; and while $k[=Exp(-\beta E_b)]$ is the stiffness of the chain. The weight factors are $u=Exp(2\beta E_s)$, $v=Exp(3\beta E_s)$ and but $w=Exp(\beta E_s)$.

## 3. The results

We calculated the grand canonical partition function of an infinitely long copolymer chain and the copolymer polymer chain is made of four monomers i. e. *a*, *t*, *c* and *g*. The partition function of the copolymer chain is written as,

$$G(g,w,k) = \frac{2((w(w(w(w(-Aw(w+1)+20B-52)+3A)+20B-52)-44B+12)-44B+12)w^2-69B-307)}{k(w^2(w(w(w(Aw^5-16(B-9)w-18B+98)-16(B-9))+8(5B-13))+8(5B-13))-3(43+13B))} \quad (2)$$

where, $A=9\sqrt{17}-49$ and $B=\sqrt{17}$

The singularity of the partition function (*i. e.* eqn. no. (2)) is used to find the probability of the formation of an infinitely long self-avoiding semi-flexible copolymer chain. It has been shown that the bonding energy ($E_s$) of the self-avoiding chain is independent of the bending energy of the copolymer chain. This nature of $E_s$ Versus $E_b$ has been shown in figure no. 2.

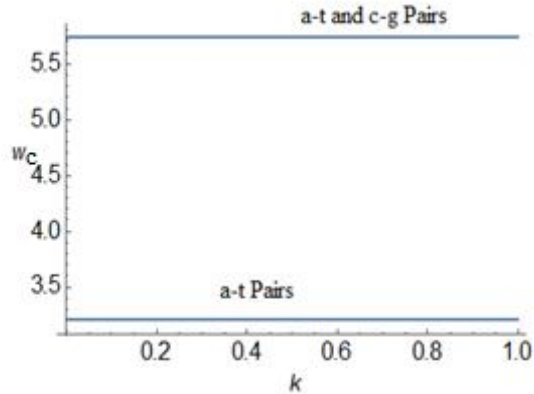

**Figure No. 2:** It is shown in this figure that below ($w_c=$) 3.2 there are neither *a-t* nor any *c-g* bonding in the copolymer chain, however above 5.73 there are *a-t* and *c-g* bonding possible in the copolymer chain conformations. Thus, in between the value of $w_c$ 3.2 to 5.73 there is only *a-t* bonding probable in the self-avoiding copolymer chain conformations.

We calculated an average number of *a-t* bonds in the copolymer chain using following equation,

$$<N_u> = \frac{\partial Log(G)}{\partial Log(u)} \bigg/ \frac{\partial Log(G)}{\partial Log(g)} \quad (3)$$

we calculated an average number of c-g bonds in the semi-flexible copolymer chain using following relation,

$$< N_v > = \frac{\partial Log(G)}{\partial Log(v)} \Big/ \frac{\partial Log(G)}{\partial Log(g)} \qquad (4)$$

The average number of *c-g* to *a-t* bonding pairs can be calculated using following relation;

$$r = \frac{<N_v>}{<N_u>} \qquad (5)$$

We have shown the values of $<N_u>$ and $<N_v>$ in the figure no. (3) and (4) respectively.

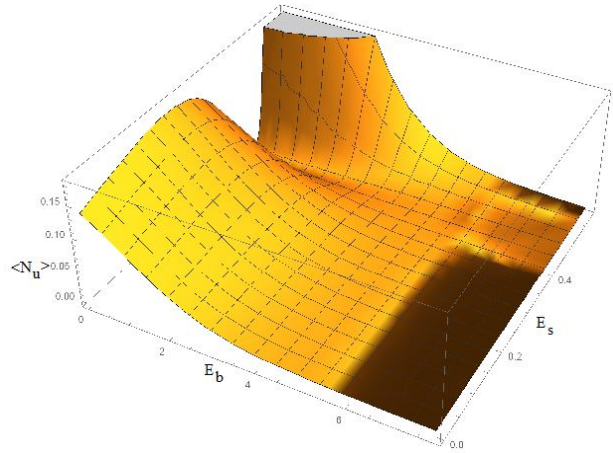

**Figure No. 3:** This figure is used to show the average number *a-t* bonds in an infinitely long semi-flexible copolymer chain. It is shown that there is one peak (at $E_s = 0.58$) in the distribution of *a-t* bonds in the semi-flexible self-avoiding copolymer chain.

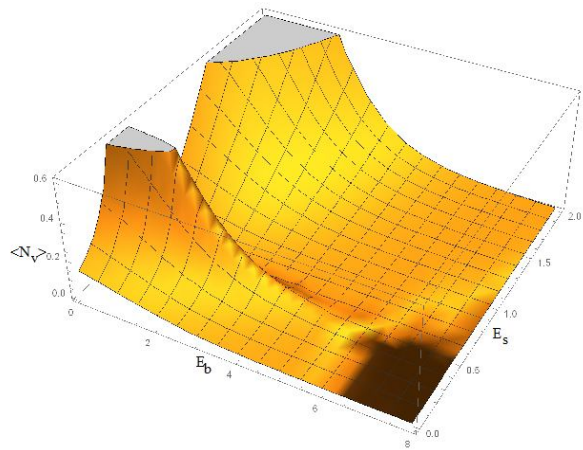

**Figure No. 4:** We have shown the average number of the *c-g* pairs in an infinitely long self-avoiding copolymer chain for its various values of the bedning energy and the bodning energy. It is seen from this figure that graph has a peak in the c-g pairs number distribution and peak is located at $E_s = 1.74$ in addition to a peak at $E_s = .58$.

## 4. The summary and the conclusions

We model an infinitely long semi-flexible copolymer chain as a partially directed self-avoiding walk [2-5] to mimic the conformations of the copolymer chain. All walks of the chain starts from a point *O* and the walker is allowed to take steps along +*x*, ±*y* and +*z* directions on a cubic lattice to mimic the conformations of the copolymer chain in three dimensions. The Boltzmann weights *u\*g* and *v\*g* are the effective Boltzmann weight on the formation of an *a-t* bond and a *c-g* bond, respectively. We have obtained the critical value of the step fugacity [$g_c(k)$] of a partially directed self-avoiding semi-flexible homo-polymer chain [5], and this value has been used to calculate the grand canonical partition function of an infinitely long copolymer chain, and its [$g_c(k)$] value is shown as follows;

$$g_c(k) = \frac{\sqrt{17}k - k - 2}{2(4k^2 - k - 1)} \tag{6}$$

Thus, from the singularity of the partition function of the copolymer chain, we estimate the probability of formation of an infinitely long copolymer chain and it has been found from our calculations that below ($w_c$=) 3.2, a copolymer chain has been polymerized without any *a-t* or the *c-g* bonds. However, above *($w_c$=)* 5.73, we have both *a-t* and *c-g* bonds which are formed in the self-avoiding semi-flexible copolymer chain. While for the $w_c$ values in between 3.2 to 5.73, we have only *a-t* bonds which are formed in the self-avoiding copolymer chain. It is seen from our analytical calculations that the bending energy has no role to play in the formation of a self-avoiding semi-flexible copolymer chain; therefore, $w_c$ ($u_c$ and $v_c$) is independent of the stiffness of the chain. However, $w_c$ ($u_c$ and $v_c$) is function of stiffness of the chain for the Gaussian copolymer chain [10].

We speculate that such distinction in the possibility of formation of a self-avoiding copolymer chain than the Gaussian copolymer chain where the stiffness of the chain has no role to play for the formation of a self-avoiding copolymer chain may be due to fact that the bending energy and bonding energy were compensated while the formation of the self-avoiding copolymer but such compensation of these energy is not possible in the case the formation of the Gaussian copolymer chain. It is due to fact that the Gaussian chain has very large entropy than the self-avoiding copolymer polymer chain.

It is seen from figure no. (3) and (4) that the average number of the *a-t* and the *c-g* bonds in the copolymer chain has a peak. This is shown from figure nos. (3) and (4) that a large number of the *a-t* and the *c-g* bonds are formed in the case of a self-avoiding flexible copolymer chain that a self-avoiding stiff copolymer chain. In other words, the stiffness of the chain restricts bending of the chain and hence in the proposed model system stiff copolymer chain formation is less probable than a self-avoiding flexible copolymer chain.

We have also shown in the figure no. 5 the ratio of *a-t* to that of the *c-g* bonds in the copolymer chain for given values of the bending energy and the bonding energy of a self-avoiding copolymer chain. The bonds ratio (*r*) varies with bending energy and bonding energy of a self-avoiding copolymer chain. However, the persistent length [5, 9-12] of the chain is shown in figure no. 6. The persistent length of the chain is the function of the bending energy of the chain.

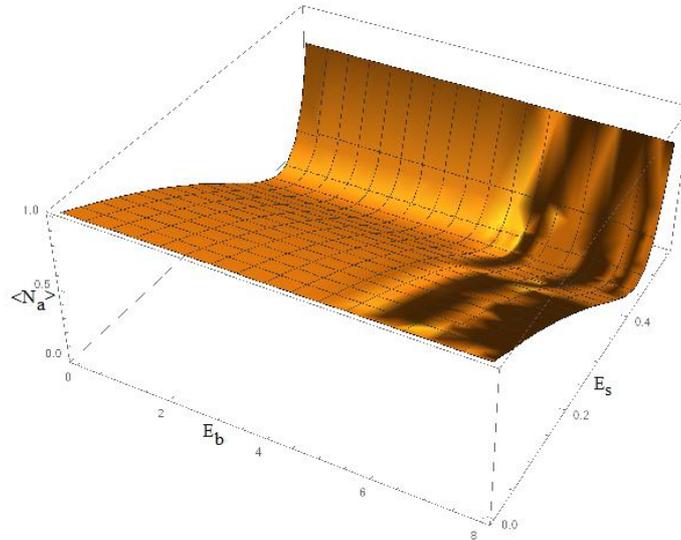

**Figure No. 5:** The ratio of average number of the *a-t* to that the *c-g* bonds is shown in this figure. The ratio diverges at the $E_s$=.58.

The possibility of the formation of a self-avoiding copolymer chain is function of the bonding energy of the copolymer chain but is independent of stiffness of the chain. The possibility of the formation of a self-avoiding copolymer chain is also a function of the monomer sequence; and it has been found that in the case of chosen sequence of the different monomers, whenever there is possibility of the formation of both *a-t* and *c-g* bond pairs in the copolymer chain, the proposed phase diagram (i. e. figure no. 2) will remain the same. However, if there is only possibility of the formation of either only *a-t* or merely *c-g* bonds in the copolymer chain the phase boundary will change. We have also found that for all the possible sequences in the monomers for the formation of a self-avoiding semi-flexible copolymer chain, the possibility of the formation of a self-avoiding semi-flexible copolymer chain will remain independent of the stiffness of the chain.

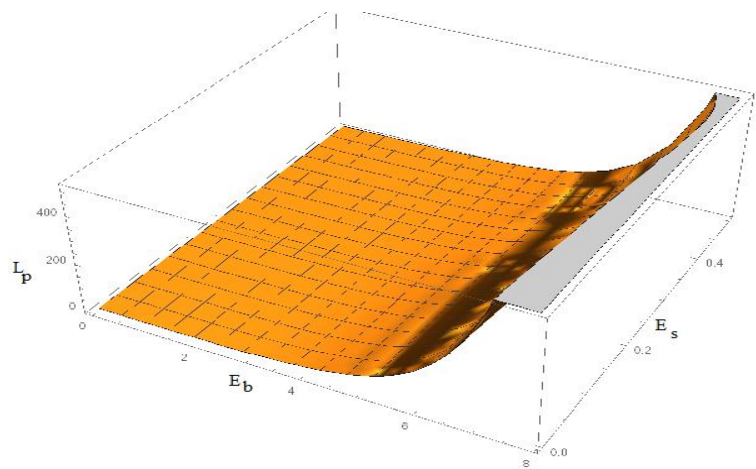

**Figure No. 6:** The persistent length of a self-avoiding semi-felxible copolymer chain is shown in this figure. It is seen that the persistent length is fucntion of the bending energy of the copolymer chain.

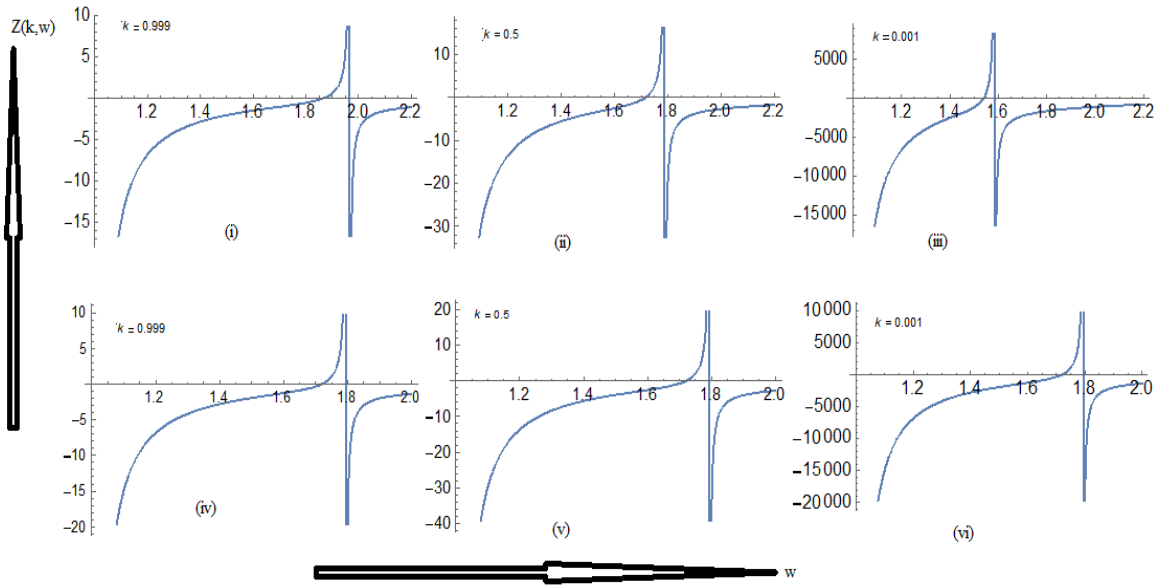

**Figure No. 7:** We have shown the nature of the grand canonical partition function [$G(k, w)$] of the Gaussian semi-flexible copolymer polymer chain in the figure nos. 7 (i-iii) and we have also shown the partition function of the self-avoiding semi-flexible copolymer chain in the figure nos. 7(iv-vi) to compare the transitions which are occurring in the both the copolymer chains. It is clear from this figure that in the case of the Gaussian copolymer chain the transition is function of the stiffness of the copolymer chain while in the case of a self-avoiding semi-flexible copolymer chain the transition is independent of the stiffness of the copolymer chain. However, in the both of the cases *i. e.* in the case of the Gaussian copolymer chain and the self-avoiding semi-flexible copolymer chains the transition which is occurring at higher value of bonding energy of the chain is a strong transition than the transition that is occurring at lower value of the bonding energy of the monomers of the copolymer chain.

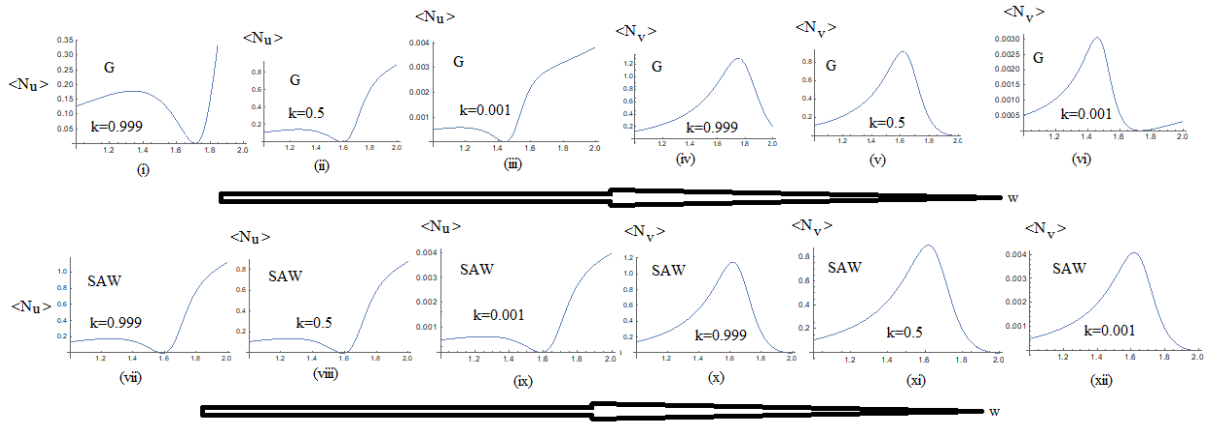

**Figure No. 8:** We have shown the average number of *a-t* and *c-g* bonds for the chosen values of the stiffness ($k$) of the Gaussian copolymer chain in the Figure no. 8(i-vi) and we have also shown the average number of *a-t* and *c-g* bonds in the self-avoiding copolymer

chain to distinguish the transitions which are occurring in the copolymer chain at low and also at the high value of the bonding energy for the Gaussian and self-avoiding copolymer chains.

We have shown a comparision on the phase transitions which have been justified on the basis of the singularity of the grand canonical partition function of a Gaussian semi-flexible copolymer chain and a self-avoiding semi-flexible copolymer chain, and we have showed the nature of the partition fucntion in the figure no. 7. It is shown further that the nature of the transition has distinction when we consider stiffness of the chain; however, in the case of transition at the lower value of the bonding energy of the monomers; the bonding of different monomers is a weak transition than the formation of a copolymer chain, which is a strong transition at the higher value of the bonding energy of the copolymer chain. The nature of transitions at the lower and higher values of the bonding energy of a self-avoiding copolymer chain and the Gaussian copolymer chains have also been shown in the figure no. 8.